# The danger of pseudo science in Informetrics

*Jos AE Spaan,*

*Professor Medical Physics at the Academic Medical Center, University of Amsterdam.*

In their recent paper 'Caveats for the journal and field normalizations in the CWTS ("Leiden") evaluations of research performance' Opthof and Leydesdorff (2010) point out that it is important to first 'multiply and divide' before 'adding and subtract' instead of the other way round, as promoted by the CWTS. In their reply van Raan et al (2010) state that: '… the order of operations argument is irrelevant in the choice between the two normalization methods'.

The statement of van Raan et al is remarkable. Is it allowed to arbitrarily choose a method of normalization or is there logic behind the definition of normalization of group related variables? The problem is that the scientific notions under consideration are vague and difficult to measure, and that the data analysis is not hypothesis driven but merely descriptive. In a scientific approach we therefore often fall back on 'analogues' so we can compare or model a certain problem with poorly defined variables and a poorly defined structure to processes where these uncertainties are absent. For example, we can understand the flow of blood in the circulation by describing it is as an electrical current so that concepts from electrical engineering as compliance, resistance, inductance, pressure, power and more can be applied in an analogues way. Many other examples can be provided.

The relevance of the order or operations becomes clear by a simple analogon. Let us take the analogy between the Citations and JCS on the one hand with weight, W, and length, L, of a group of people on the other hand. Clearly, one may calculate the average weight and the average length of the group of subjects. However, it is well appreciated that in general longer persons will have a larger weight and hence the judgment on the general health of the group requires normalization. The Quetelet index has been developed where weight is divided by the length square but for the sake of argument we will consider here the simple ratio between weight and length, W/L. According to the 'CWTS order of operations' we would be allowed to use the ratio between average weight and average length to judge the health condition of our group of subjects. According to the rules of order of operations and common sense, we have to calculate normalization on a per person basis and then average the normalized index per person over the group. In any case, would the data of table 2 in the paper of Opthof and Leydesdorff apply to length and weight of the group, the normalized weight according to the CWTS rule would be 30% too low while the Opthof and Leydesdorff index would indicate a healthy group. A decision to supplement food to the group on the basis of the CWTS index in order to reach a healthy normalized weight would lead to obesity. The order of mathematical operations therefore does matter.

One can think of many other examples where the order of operations does matter. For example, there will be no physicist or engineer that, when asked to report an average electrical power consumption of a group of houses, to first average all currents and voltages at the main line into each house and then take the product of the two averages. No, the power consumption will be calculated per house (for which voltage and current for each house are multiplied) and then averaged. I challenge van Raan and coworkers to come with one example in a field of science where the measuring units have physical or financial meaning and where it would be allowed to first take averages and then divide or multiply the averages. My contention is that there is none.

Why would the emerging field of Informetrics be different and so unique that normal mathematical rules would not be applicable? These rules are not just arbitrary but underlie scientific principles which are clear in all scientific areas where quantities can objectively be measured: think of conservation of mass, energy, impulse and so on. May be the difference is that Scientific quality is not an objective measure and hence that the analysis method of Informetrics does not need to follow the normal scientific rules. But then one may easily conclude that we have to deal with Pseudo Science.

The further defense of van Raan by statistical comparison of the two methods is not to the point. A statistical correlation is not to be used as a scientific argument on the application of general methodological concepts to science including Informetrics. The principle of calculation is right or wrong and the CWTS way of normalizing for a group is just wrong. The scatter plot of correct versus CWTS indices in the reply letter of Raan at al. is only useful in identifying those scientists that were given an advantage in the AMC system and those who could team up for a class action lawsuit because they are victim of the arbitrary choice of CWTS normalization.

The practical problem is that CWTS produces evaluations of performance of individual scientists to institutions as the Academic Medical Center of the University of Amsterdam, AMC. These evaluations play a crucial role in reorganizations of these institutions where especially interdisciplinary sciences are suffering (Spaan 2009) . The CWTS has therefore created for itself a huge responsibility. In the Dutch public press, van Raan has claimed that it is clear that the CWTS analysis should not be used to judge individual scientists because of the sample size of publications. However, when asked, this information is provided by the CWTS to these institutions and consequently used on an individual scale.

Real science has always suffered from pseudo science but now this has risen to a different level and becomes a real threat to science when pseudo science is applied by our managers as an instrument to decide on the scientific survival of excellent individual scientists and fields of research.

References:

Opthof, T., & Leydesdorff, L. (2010, in press). Caveats for the journal and field normalizations in the CWTS (―Leiden‖) evaluations of research performance. *Journal of Informetrics* .

van Raan A.F.J. , van Leeuwen T.N , Visser M.S. , van Eck N.J. , Waltman L (2010, in press) Rivals for the crown: Reply to Opthof and Leydesdorff. *Journal of Informetrics*

Spaan J.A. (2009) Biomedical engineering and bibliometric indices for scientific quality. Med Biol Eng Comput (2009) 47:1219–1220